\documentclass[conference,compsoc]{IEEEtran}
\pdfoutput=1 
\IEEEoverridecommandlockouts

\usepackage{cite}
\usepackage{booktabs}
\usepackage{multirow}
\usepackage{xcolor}
\usepackage{amsmath}
\usepackage{amssymb}
\usepackage{subcaption}
\usepackage{enumitem}
\usepackage{algorithm}
\usepackage{algpseudocode}
\usepackage{graphicx}
\usepackage{xurl}
\usepackage{tabularx}
\newsavebox{\cmdbox}

\makeatletter
\renewcommand\paragraph{\@startsection{paragraph}{4}{\z@}%
  {1.4ex \@plus 1ex \@minus .2ex}{-0.6em}%
  {\normalfont\normalsize\bfseries}}
\def\@IEEEsectpunct{\ }
\makeatother

\begin{document}

\title{When Safe Skills Collide: Measuring Compositional Risk in Agent Skill Ecosystems}

\author{%
\IEEEauthorblockN{Su Wang\textsuperscript{1,*}, Pin Qian\textsuperscript{1,*}, Yihang Chen\textsuperscript{2,*}, Junxian You\textsuperscript{3}, Xiaoyuan Wang\textsuperscript{1},\\
Xiaochong Jiang\textsuperscript{4}, Lifei Liu\textsuperscript{4}, Haoran Yu\textsuperscript{4}, Jingzhou Xu\textsuperscript{5}}
\IEEEauthorblockA{%
\textsuperscript{1}Carnegie Mellon University \quad
\textsuperscript{2}Georgia Institute of Technology\\
\textsuperscript{3}University of Glasgow \quad
\textsuperscript{4}Independent Researcher\\
\textsuperscript{5}Corespeed Inc.}
\thanks{\textsuperscript{*}These authors contributed equally.}}

\maketitle

\begin{abstract}
LLM agents increasingly rely on community-contributed skills that expand an agent's operational capability set.
We study a core safety problem in agentic AI systems: whether individually safe skills can compose into unsafe installed skill sets.
We present \textsc{SkillReact}, a compositional security measurement framework with three components: a deterministic static-composition benchmark, a two-rater LLM-assisted human-adjudication pipeline, and an action-based exploitability harness.
On 1{,}520 ClawHub skills, 651 pass individual inspection and form $211{,}575$ pairs; the benchmark flags $22.25\%$ of these as structural candidates.
We treat this raw rate as a recall-oriented scanner ceiling and calibrate it against human judgment: in a pattern-stratified audit, roughly one in five flagged pair-pattern hits survives as a real compositional risk (population-weighted validity $\boldsymbol{18.2\%}$, our headline result), implying about $14$K genuine risk memberships in a single registry that per-skill scanning misses by construction, since every pair is individually safe.
An action-based harness then probes when these candidates become model-issued tool calls, and finds realization gated by host-model disposition: on an anchor-conditioned dropper subset, Haiku-4-5 issues the dropper-stage tool call on all $39$ direct-prompt trials ($36$ of them the full download-then-execute chain, $3$ download-only), Opus-4-7 stops at the download, and Sonnet-4-6 refuses outright.
A control that holds the request fixed and varies only the installed skills finds compliance \emph{highest with no skills installed}: a composition fixes which capabilities are reachable, while the host model decides whether to use them.
Together these motivate install-time compositional checks and capability isolation as complements to per-skill scanning.
\end{abstract}


\medskip

\section{Introduction}
\label{sec:intro}

Agent skill ecosystems have grown rapidly: large language model (LLM) coding assistants such as Claude Code, Cursor, and Windsurf load community-contributed skills---bundles of instructions, references, and automation scripts---that grant the host agent operational capabilities such as file access, network egress, credential handling, and process spawning~\cite{anthropicskillsdocs,anthropicskills2025}.
The dominant security response so far has been to scan one skill at a time: per-skill malicious-skill detectors~\cite{maliciousskills2026}, vulnerability scanners~\cite{skillscan2026,skillsieve2026,skillfortify2026}, and capability classifiers~\cite{toxicskills2026,connor2026}.

Deployed agents, however, do not run one skill at a time.
A user with a file-read--only research assistant and a separate network-only API helper installs both into the same unified runtime; the agent inherits the union of their capabilities.
Each skill is individually safe under a capability policy that forbids exfiltration channels, yet their union satisfies the file-read $\wedge$ network-outbound conjunction that defines the policy violation.
The relevant security object is therefore not the package; it is the installed skill set available to the agent in a single session.

\begin{figure*}[t]
  \centering
  \includegraphics[width=\textwidth]{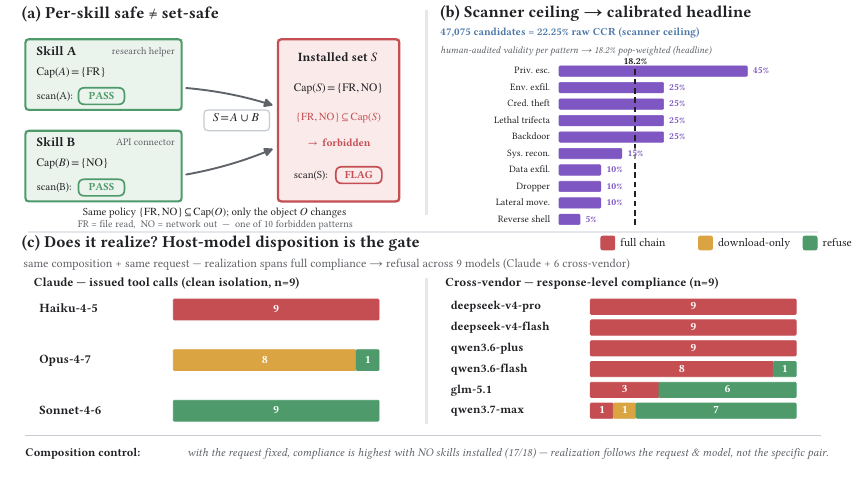}

  \caption{SkillReact in one view. \textbf{(a)~Per-skill safe $\neq$ set-safe:} the same capability policy---flag an object $O$ iff $\{$FR,\,NO$\}\subseteq\mathrm{Cap}(O)$ (file read $\wedge$ network out, one of the 10 forbidden patterns)---clears each skill individually ($\mathrm{Cap}(A)=\{$FR$\}$, $\mathrm{Cap}(B)=\{$NO$\}$: both PASS), yet flags their co-installed union ($\mathrm{Cap}(S)=\{$FR,\,NO$\}$, so $\{$FR,\,NO$\}\subseteq\mathrm{Cap}(S)$: FLAG); only the security object changes (skill $\to$ installed set), not the rule. \textbf{(b)~How common (calibrated):} the deterministic scanner flags $47{,}075$ of $211{,}575$ individually-safe pairs ($22.25\%$ raw CCR, a recall-oriented scanner ceiling); bars give the human-audited validity of each of the 10 forbidden patterns ($5$--$45\%$, from $200$ audited units), whose population-weighted mean is the $18.2\%$ pair-pattern validity rate (\emph{the headline}, dashed line). Under an added co-validation assumption this projects to a $4.05\%$ pair-level point estimate; the only assumption-free statement is the ordering (unique-pair rate $\le$ membership rate), a projected $\le\!6.78\%$ cap. \textbf{(c)~Whether it realizes:} the same composition and request yield a realization gradient tracking host-model disposition across $9$ models, at two \emph{non-pooled} levels (Table~\ref{tab:disposition_gradient})---Claude issued tool calls (clean isolation) and cross-vendor response-level compliance---spanning full download-then-execute chains to near-total refusal. A composition control finds compliance \emph{highest with no skills installed} ($17/18$), so the composition supplies the static capability surface while host-model disposition---not the specific pair---gates realization.}
  \label{fig:overview}
\end{figure*}

This compositional blind spot is exactly what per-skill scanning misses by construction.
Spera~\cite{spera2026formal} recently established a non-compositionality result for capability hypergraphs: agents whose individual capability sets each fail to reach any forbidden capability can, when composed, reach forbidden capabilities through conjunctive composition rules.
That result is a structural existence proof on abstract graphs.
It does not predict how often the failure mode arises in real registries, what shape it takes, or when the static condition turns into an actual model-issued tool-call attempt.
We do not claim Spera's theorem proves our empirical result; we use it only to fix the unit of analysis.

We therefore ask:
\begin{quote}
\emph{How common are structural compositional-risk candidates in real skill ecosystems, and when do selected candidates realize as model-issued tool-call attempts?}
\end{quote}

We answer with \textsc{SkillReact}, a compositional security \emph{measurement framework}.
The framework comprises (i) a deterministic static-composition benchmark over a snapshot of ClawHub, (ii) a completed LLM-assisted human-adjudication pipeline that calibrates the static flags, and (iii) an action-based exploitability harness that records when models actually invoke tools toward the forbidden goal.
We are explicit that the static side measures a recall-oriented upper bound (a \emph{candidate} rate, not an exploit prevalence rate) that the human-adjudication pipeline calibrates, and that the action-based side reports model- and runtime-conditional realization, not registry-wide exploit prevalence.

Our contributions are:
\begin{enumerate}[leftmargin=*]
  \item \textbf{Compositional risk formulation.} We separate individually-flagged skills, capability-union flags, calibrated structural candidates, and model-issued tool-call attempts, giving registries an auditable hierarchy between per-skill scanning and full dynamic exploitation.
  \item \textbf{Large-scale ecosystem measurement, calibrated.} On 1{,}520 real ClawHub skills, the static benchmark evaluates 211{,}575 individually-safe pairs against 10 forbidden patterns and flags 47{,}075 candidates; we report the raw \emph{Pairwise Capability-Union Rate} ($22.25\%$, 95\% CI $[19.98\%, 24.61\%]$) only as a recall-oriented scanner ceiling. Our load-bearing number is its calibration at the (pair, pattern) annotation level via a 200-unit pattern-stratified gold set---two LLMs help mark each unit and a human audits/corrects---yielding a population-weighted pair-pattern validity rate $\hat p^{\mathrm{pp}}=18.2\%$ (95\% CI $[11.3\%, 27.4\%]$): even after human audit, roughly one in five flagged pair-pattern hits is a real compositional risk that per-skill scanning misses by construction (every pair is individually safe). This corresponds to an estimated $\approx\!14.4$K valid pair-pattern risk memberships in a single registry; under a co-validation assumption it projects to a pair-level point estimate of $4.05\%$ ($\approx\!8.6$K unique risky pairs), and since valid pairs cannot outnumber valid memberships, the same estimate caps the unique-pair rate at a projected $6.78\%$ (§\ref{sec:calibrated_ccr}).
  \item \textbf{LLM-assisted adjudication protocol.} Two independent raters (Claude Sonnet 4.6 + Codex gpt-5.5) help mark every (pair, pattern) unit; a stdlib-only UI surfaces both LLM markings to a human reviewer who audits and corrects them; a stratified estimator aggregates the audited labels into the calibrated rate.
  \item \textbf{Reliability and upper-bound diagnostics.} Per-pair evidence traces, fixed-pool pattern ablation, capability-level $\Delta$CCR, negation-window analysis, inter-rater LLM agreement, and human override rate jointly expose where the static bound is loose and which capabilities are load-bearing.
  \item \textbf{Action-based exploitability harness, with composition controls.} Selected dropper candidates anchored on a single popular skill realize as model-issued dropper-stage tool-call chains under Haiku-4-5 ($39/39$ direct-prompt trials: $36/39$ full download-then-execute, $3/39$ download-only), while Sonnet-4-6 refuses and the strongest Anthropic model we evaluated (Opus-4-7) stops at download-only; a control that varies only the installed skills shows realization tracks model disposition to the explicit request, not the specific pair.
  \item \textbf{Install-time checker and design implications.} A two-mode compositional checker reports set-level violations and pair-level evidence in sub-millisecond time per install, bounded explicitly by our 10-pattern taxonomy.
\end{enumerate}

The contribution is a security \emph{measurement}: our headline is the human-calibrated validity of those candidates (and the model-disposition gradient that gates their realization), with the raw $22.25\%$ reported only as the scanner's recall ceiling.

\section{Background and Related Work}
\label{sec:background}

\subsection{Agent Skill Ecosystems}

An agent skill is a self-contained extension that augments an LLM coding assistant with new capabilities.
Skills typically consist of a manifest file (\texttt{SKILL.md}) describing purpose and usage, plus optional code files that implement tool integrations.
Skills are distributed through registries such as ClawHub~\cite{clawhub}.
When a user installs multiple skills, they operate inside the agent's unified execution context: the agent can invoke any installed skill's capabilities during a session, effectively inheriting the \emph{union} of installed capabilities.
This design maximizes utility but creates a security surface that scales combinatorially with set size.

\subsection{Capability Policy}

We model each skill as possessing a subset of eight capabilities: \emph{file read}, \emph{file write}, \emph{network outbound}, \emph{network inbound}, \emph{environment read}, \emph{process spawn}, \emph{credential access}, and \emph{system info}.
A skill is \textbf{individually unsafe under our capability policy} (or \emph{individually flagged}) if its capability profile already satisfies one of the 10 forbidden patterns of §\ref{sec:patterns}.
We use ``individually safe'' and ``individually unsafe'' strictly in this taxonomy-relative sense.
A skill that is individually safe under our policy is not necessarily benign, vulnerability-free, or safe under all possible policies; it is a skill whose own capability profile, in isolation, does not satisfy any of our 10 forbidden conjunctions.
Likewise, ``individually flagged'' is a scanner outcome, not a claim of malicious intent.

\subsection{Threat Model and Runtime Assumption}
\label{sec:threat_model}

We model a \textbf{unified, non-isolated agent runtime} in which the host agent can access all installed skills inside a single session.
This is a common worst-case abstraction; it is not a claim that every deployed runtime lacks isolation.
Runtime permissioning, skill sandboxing, network allowlists, capability scopes, and per-skill confirmation prompts can all reduce realization risk in practice, and we treat these as complementary defenses (§\ref{sec:design}).
Our measurement targets the structural blind spot itself; whether and how a given runtime closes it is a separate engineering question.

\subsection{Three Levels of Evidence}

Our measurement intentionally separates three levels of evidence about compositional risk:
\begin{itemize}[leftmargin=*]
  \item \emph{Individually flagged skill}: a single skill whose capability profile already satisfies a forbidden pattern.
  \item \emph{Structural compositional-risk candidate}: a pair of skills, each individually safe under our policy, whose capability \emph{union} satisfies a forbidden pattern. This is the primary object counted by the static benchmark.
  \item \emph{Model-issued tool-call attempt}: the host agent, exposed to the relevant skills, actually invokes tools toward the forbidden goal. Whether this happens depends on host model, prompt, and runtime context, and is measured by the action-based harness.
\end{itemize}

\subsection{Related Work}
\label{sec:related}

\paragraph{Per-skill skill security.}
SkillScan~\cite{skillscan2026}, ToxicSkills~\cite{toxicskills2026}, SkillSieve~\cite{skillsieve2026}, SkillFortify~\cite{skillfortify2026}, Connor~\cite{connor2026}, and MindGuard~\cite{mindguard2025} measure or detect risk one skill at a time; \cite{maliciousskills2026,credentialleak2026,maloyan2601} characterize malicious or vulnerable single skills; ToolHijacker~\cite{toolhijacker2026}, BadSkill~\cite{badskill2026}, DDIPE~\cite{ddipe2026}, and SkillJect~\cite{skillject2026} target individual skills on the attack side. Our contribution is orthogonal: it changes the unit of analysis from one skill to an installed skill set.

\paragraph{Compositional security lineage.}
Spera~\cite{spera2026formal} formalized non-compositionality of safety on capability hypergraphs; McCullough~\cite{mccullough1988} showed non-interference is not preserved under hookup composition; classical information-flow~\cite{denning1976lattice} and least-privilege~\cite{saltzer1975protection} motivate the same lens. We adopt Spera's capability-hypergraph framing to fix the unit of analysis. On Android, COVERT~\cite{covert2015} introduced compositional inter-app permission-leakage analysis, and MR-Droid~\cite{mrdroid2017} later scaled inter-app communication risk analysis to $11{,}996$ apps and $13$M app pairs; we adopt that pairwise-composition pipeline shape and inherit its limitations (capability-extraction noise, taxonomy-bounded completeness). The substrate change matters: where Android's execution engine is a deterministic OS that mediates IPC by manifest declarations, an agent skill's engine is an LLM whose compliance varies with prompt, anchor identity, and model disposition---our Haiku-issues-the-chain / Sonnet-refuses / Opus-4-7-downloads-only split (§\ref{sec:exploitability}) makes \emph{host-model disposition} a first-class quantity for skill registries with no analogue in COVERT-style analysis.

\paragraph{Agent and runtime defenses.}
TOP-R~\cite{topr2025} studied multi-tool privacy leaks at the tool-call level; AgentDojo~\cite{agentdojo2024}, ASB~\cite{agentsmith2024}, and InjecAgent~\cite{injecagent2024} benchmark per-tool robustness against prompt injection. Runtime defenses~\cite{camel2025,llamafirewall2025,agrail2025,agentspec2026,shi2026progentsecuringaiagents,microsoftagentgov2026} provide isolation, filtering, or guardrail infrastructure, including reasoning-based safeguards that self-reflect on adversarial prompts~\cite{lin2026reflect}; OWASP Agentic Skills Top 10~\cite{owasptop10skills} and OpenClaw~\cite{clawvulntax2026} provide standardized taxonomies, and a runtime supply-chain taxonomy maps data- and tool-supply-chain attack vectors across the agentic stack~\cite{jiang2026soktaxonomyattackvectors}. Prior work improves per-skill or runtime security, or studies how agents learn to invoke tools~\cite{xu2026learningusetoolsjust} and how that tool use is supervised~\cite{jiang2026scribe}; \textsc{SkillReact} changes the unit of analysis to the installed skill set with a measurement framework, install-time check, and action-based harness.

\section{The SkillReact Measurement Framework}
\label{sec:framework}

\textsc{SkillReact} is a security measurement framework, not just a benchmark.
It comprises three components: a deterministic static-composition benchmark, a completed LLM-assisted human-adjudication pipeline, and an action-based exploitability harness.

\paragraph{Static-Composition Benchmark (deterministic).}
The static benchmark freezes the ClawHub corpus snapshot, the capability schema, the regex catalog, and the 10 forbidden patterns.
It recomputes capability profiles, individually-safe pairs, structural candidates, and per-pair evidence traces; the entire pipeline is deterministic and version-controlled, so every downstream number recomputes identically from the frozen snapshot.
This component is the source of the raw CCR, which we report as a recall-oriented scanner ceiling; the human-adjudication pipeline below calibrates it into the headline rate.

\paragraph{Action-Based Exploitability Harness (model/runtime sensitive).}
The harness records whether a host model, when shown skill artifacts under a stated runtime policy, invokes tools (e.g., \texttt{Bash}, \texttt{Read}) toward the forbidden goal.
It contains two named protocols: \emph{Capability-Sketch Trials} present abstract capability pairs, and \emph{Skill-Artifact Trials} present real \texttt{SKILL.md} pairs.
Because results depend on host model, prompt strategy, and runtime, we treat this component as conditional realization evidence, not a deterministic prevalence estimate.

\paragraph{Human-Adjudication Pipeline.}
We run a stratified 200 pair-pattern gold set through two LLM raters and a human auditor, with the matched evidence shown at adjudication.
This pipeline is \emph{completed}: its human-audited per-pattern validity rates are the source of the calibrated CCR (§\ref{sec:calibrated_ccr}).

The framework's primary outputs are: the raw CCR over structural candidates and per-pattern candidate counts; the human-calibrated pair-pattern validity rate (the headline measurement); per-pair evidence traces and capability-level diagnostics; and per-pair logs of tool-call attempts.

\section{Methodology}
\label{sec:method}

\subsection{Skill Collection}
\label{sec:collection}

We collected 1{,}520 skills from ClawHub by attempting retrieval of all 5{,}128 registry entries exposed by our crawler at collection time; the remainder failed during download and were not selected on any security-relevant criterion.
Our collection log recorded $286$ of these download failures. Re-attempting those $286$ against the live registry, $269$ ($94.1\%$) are in fact retrievable; the remaining $17$ were not recovered (no manifest found under any author/skill split we tried, which is not the same as verified absence): the \emph{recorded} failures were dominated by two causes unrelated to skill content---a source path that has since been removed, and an author/skill slug-splitting heuristic that mis-resolved hyphenated author names---both now corrected.
This shows the recorded attrition was largely a tooling artifact, but it is a limited check: $286$ is only the logged subset of the full attrition, so we do not claim it characterizes the unlogged remainder, nor that the failed skills' capability profiles match the analyzed corpus.
We retain the $1{,}520$-skill snapshot as the frozen unit of analysis and treat capability-profile independence of attrition as an \emph{assumption} rather than a measured property.
Prior work reports larger ClawHub-scale evaluations, so our snapshot is a substantial but incomplete slice of the ecosystem.
The successful downloads span productivity tools, development utilities, system automation, communication integrations, and data analysis packages.
This is, to our knowledge, the largest pairwise compositional analysis of agent skills reported to date; per-skill datasets such as ToxicSkills~\cite{toxicskills2026} contain more individual skills but do not perform pairwise capability-union analysis.
Skills were collected without pre-filtering for security properties.
For each skill we retrieved the complete \texttt{SKILL.md} manifest and all associated code files (Python, JavaScript/TypeScript, and shell scripts).

\subsection{Capability Extraction}
\label{sec:extraction}

The main pipeline uses a deterministic, recall-oriented regex extractor.
Capability labels produced by the extractor are not ground truth; they are reproducible scanner outputs, and the headline metric inherits this caveat by design.

\paragraph{Static Analysis.}
For each skill we scan \texttt{SKILL.md} and bundled code with a fixed regex catalog covering the 8 capabilities of §\ref{sec:background}; a capability fires if either a code pattern or a natural-language pattern over the skill prose matches.
Each fire is recorded as an \emph{evidence trace}: the matched file, the matched regex identifier, and a $\pm$40-character snippet.
The full pattern list is version-controlled and fixed across all downstream numbers.

\paragraph{Capability-Level LLM Cross-Check (auxiliary diagnostic).}
On a frozen 50-skill gold subset we obtained a second-rater capability labeling using Claude Sonnet 4.6 (\texttt{claude-sonnet-4-6}, accessed 2026-04-19) with a fixed prompt.
Agreement against the regex labels at the per-skill capability level is reported as Cohen's $\kappa$ in §\ref{sec:diagnostics}.
This capability-level LLM cross-check is auxiliary; production capability profiles remain regex-only and bit-for-bit reproducible.
A separate, much stronger calibration runs at the (pair, pattern) level via the LLM-assisted human-adjudication protocol of §\ref{sec:human_protocol}.

\subsection{LLM-Assisted Human-Adjudication Protocol}
\label{sec:human_protocol}

The capability-level $\kappa$ above is a noise diagnostic, not a validity rate over flagged pairs.
To obtain a validity rate at the level of the headline metric we run a 200-unit stratified gold set through a two-rater LLM-assisted human-adjudication pipeline.

\paragraph{Gold sampling.}
We draw 200 pair-pattern annotation units from the 47{,}075 flagged candidates, stratified by the 10 forbidden patterns (20 units per pattern). Sampling is seeded and version-controlled.

\paragraph{Two LLM raters (help mark; humans audit).}
Our two-rater design instantiates the LLM-as-judge paradigm~\cite{li2025generation}, whose documented benefits and failure modes are exactly why every unit is still human-audited.
Each unit is independently marked by two LLM raters from different vendors:
(i) \emph{Rater A} = Claude Sonnet 4.6 via \texttt{claude -p --bare} with no tools and no session persistence;
(ii) \emph{Rater B} = OpenAI Codex CLI 0.128.0 invoking gpt-5.5 (accessed April 2026) via \texttt{codex exec} with \texttt{model\_reasoning\_effort=low}, \texttt{--ignore-user-config}, \texttt{--ephemeral}, and \texttt{--output-schema} bound to our shared output schema.
Because superficial prompt properties such as tone can shift model outputs~\cite{cai2025tone}, both raters share an identical prompt that includes the SKILL.md text of both skills, the regex-derived capability profile, the regex evidence traces, the target pattern definition, and a request for a structured judgment with nine fields: verdict $\in \{$VALID, FALSE\_POSITIVE, BENIGN\_BY\_CONTEXT, UNCLEAR$\}$, per-skill capability validity, target-pattern validity, risk plausibility, regex-misfire flag, confidence, an evidence-cited list, and a free-text rationale.
Output is JSON, schema-validated programmatically.

\paragraph{Stratification.}
For each unit we then assign one of five agreement strata: both raters VALID, both NOT-VALID, only-A VALID, only-B VALID, or any-UNCLEAR.
\textbf{LLM agreement is a stratum, not validation.}
Every one of the 200 units is then human-adjudicated regardless of stratum; the five strata are used for diagnostics---reporting where the two raters agree, where they disagree, and where the human verdict overturns them---not to subsample human effort.
This is the protocol that addresses the bias that two LLMs sharing training-data priors may agree wrongly on regex-artifact pairs.
Using model raters to approximate human judgment trades cost against fidelity~\cite{zhang2026performance}, which is why the LLM markings here are triage signals rather than final labels.

\paragraph{Adjudication UI.}
A stdlib-only Python HTTP UI surfaces, for each unit: the (pair, target pattern) header; both skills' SKILL.md; per-pattern regex evidence; both LLM markings (verdict, per-cap validity, rationale, evidence cited); and a 9-field human form mirroring the LLM schema.
Filters select disagreement-only and agreement-sample views; stratum and sampling-reason chips are shown on every unit card.

\paragraph{Stratified calibrated estimator (units explicit).}
With 20 units per pattern, per-pattern validity rates $\hat p_p$ come directly from human audit. The directly-measured headline quantity is the population-weighted \emph{calibrated pair-pattern validity rate} $\hat p^{\mathrm{pp}} = \sum_p (N_p^{\mathrm{pp}}/\sum_q N_q^{\mathrm{pp}})\,\hat p_p$, where $N_p^{\mathrm{pp}}$ is pattern $p$'s share of the $78{,}878$ flagged pair-pattern memberships (Table~\ref{tab:patterns}); CIs are 1{,}000-iteration per-pattern bootstrap.
$\hat p^{\mathrm{pp}}$ is at the (pair, pattern) annotation-unit level: it estimates the fraction of flagged pair-pattern memberships that are real compositional risks under human audit. Two projections follow, both derived from the \emph{same} sampled $\hat p^{\mathrm{pp}}$ and inheriting its CI. Directly, $\hat p^{\mathrm{pp}}$ applied to the $78{,}878$ flagged memberships gives an estimated $\approx\!14{,}356$ valid memberships. The assumption-free fact here is an \emph{ordering}, not a number: distinct pairs have disjoint (pair, pattern) memberships and each genuinely risky pair carries at least one valid membership, so the count of unique risky pairs cannot exceed the count of valid memberships---i.e.\ the unique-pair rate cannot exceed the membership rate. Under the point estimate this caps the unique-pair rate at $14{,}356/211{,}575 = 6.78\%$, but $6.78\%$ is itself a projection of that estimated membership count (its CI scales with $\hat p^{\mathrm{pp}}$), not a data-certain ceiling. Multiplying $\hat p^{\mathrm{pp}}$ by the raw $22.25\%$ pair rate instead yields a pair-level \emph{point estimate} of $4.05\%$ ($\approx\!8{,}567$ pairs), under the additional \emph{co-validation} assumption---that a pair's overlapping pattern hits share validity, so its membership-level and pair-level validity coincide. We deliberately do \emph{not} frame $4.05\%$ as a lower bound: because subsumed patterns (Lethal Trifecta over Data Exfiltration and Credential Theft) make genuine risks cluster into multi-pattern pairs, the true pair rate could fall below $4.05\%$ even at the estimated membership count, so the only assumption-free one-sided statement is the ordering (pair rate $\le$ membership rate), with $6.78\%$ its point-estimate value. An unbiased per-unique-pair estimate would require labelling each gold pair across \emph{all} its matching patterns, which we leave to future work. We also report a uniform-pattern variant as a sensitivity check.

\subsection{Forbidden Composition Patterns}
\label{sec:patterns}

We define 10 forbidden composition patterns by intersecting four established sources: ToxicSkills' empirical skill-threat taxonomy~\cite{toxicskills2026}, Willison's lethal-trifecta framing~\cite{willison2025trifecta}, the OWASP Agentic Skills Top 10~\cite{owasptop10skills}, and classical capability-misuse archetypes (privilege escalation, reverse shell, persistent backdoor) carried over from the Android-app composition literature~\cite{covert2015}.
Each pattern specifies a capability conjunction over a combined skill profile.
We abbreviate capabilities as FR/FW (file read/write), NO/NI (network out/in), ER (env read), PS (process spawn), CA (credential access), SI (system info).

\begin{enumerate}[leftmargin=*]
  \item \textbf{Data Exfiltration}: FR $\wedge$ NO.
  \item \textbf{Credential Theft}: CA $\wedge$ NO.
  \item \textbf{Lethal Trifecta}~\cite{willison2025trifecta}: CA $\wedge$ FR $\wedge$ NO.
  \item \textbf{Privilege Escalation}: ER $\wedge$ PS.
  \item \textbf{Reverse Shell}: NI $\wedge$ PS.
  \item \textbf{Persistent Backdoor}: FW $\wedge$ PS.
  \item \textbf{Lateral Movement}: CA $\wedge$ NO $\wedge$ PS.
  \item \textbf{Environment Exfiltration}: ER $\wedge$ NO.
  \item \textbf{System Reconnaissance Exfiltration}: SI $\wedge$ NO.
  \item \textbf{Dropper}: NO $\wedge$ FW $\wedge$ PS.
\end{enumerate}

Our Lethal Trifecta captures only the two \emph{capability} legs of Willison's framing---private-data access (instantiated as CA$\wedge$FR) and external communication (NO); the remaining leg, exposure to untrusted content, is an attack-delivery property exercised in the harness (§\ref{sec:exploitability}), not a static capability.

The patterns are \emph{not disjoint}: longer conjunctions (Lethal Trifecta, Lateral Movement, Dropper) are subsumed by shorter ones in the static taxonomy, and we report this explicitly in §\ref{sec:results}.
Severity assignments follow standard practice: patterns involving credential access or full execution chains (Dropper, Lateral Movement, Lethal Trifecta) are CRITICAL; data and environment exfiltration patterns are HIGH; reconnaissance patterns are MEDIUM.

\subsection{Pairwise Compositional Analysis}
\label{sec:pairwise}

Starting from the 651 skills that are individually safe under our policy, we form $\binom{651}{2} = 211{,}575$ pairs.
For each pair $(s_i, s_j)$ we compute the capability union $\text{Cap}(s_i) \cup \text{Cap}(s_j)$ and check whether it satisfies any forbidden pattern.
A pair is a \emph{structural compositional-risk candidate} iff (1) each skill is individually safe under our policy and (2) the union satisfies at least one forbidden pattern.

The framework's central structural metric is the \textbf{Compositional Candidate Rate} (CCR):
\[
  \mathrm{CCR} = \frac{\#\,\text{structural candidate pairs}}{\#\,\text{individually-safe pairs}}.
\]
CCR is a deterministic, recall-oriented \emph{candidate} rate.
It is conditional on (i) our 10-pattern taxonomy, (ii) our regex capability labels, and (iii) the unified-runtime threat model; it is not a realized exploit prevalence rate.

\section{Results}
\label{sec:results}

\subsection{Individual Skill Scan}

\begin{figure}[t]
  \centering
  \includegraphics[width=\columnwidth]{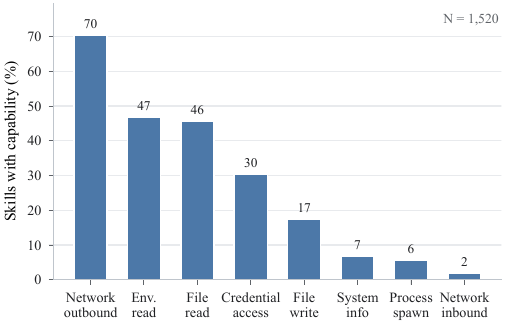}
  \caption{Distribution of capabilities across 1{,}520 analyzed skills. Network outbound (70.5\%) is near-ubiquitous; environment read (46.6\%) and file read (45.6\%) are the next most common, so the file-read\,$\wedge$\,network-outbound conjunction behind compositional data exfiltration is widely available when skills are co-installed.}
  \label{fig:cap_dist}
\end{figure}

Of the 1{,}520 skills analyzed, 869 ($57.2\%$) are individually flagged under our policy and 651 ($42.8\%$) pass individual inspection.
We call this fraction the \emph{Individual Flag Rate} (IFR).
IFR is substantially higher than prior reported per-skill vulnerability rates (e.g., SkillScan 26.1\%~\cite{skillscan2026}, ToxicSkills 36\%~\cite{toxicskills2026}).
This is a scanner diagnostic, not a vulnerability prevalence claim: our regex extractor is capability-level and recall-oriented, fires on prose mentions as well as code, and any single skill possessing both file read and network outbound is already flagged at the individual level.
Section~\ref{sec:diagnostics} (Table~\ref{tab:diagnostics}) quantifies the regex's contribution per capability.

\subsection{Compositional Candidate Rate}

\begin{figure}[t]
  \centering
  \includegraphics[width=\columnwidth]{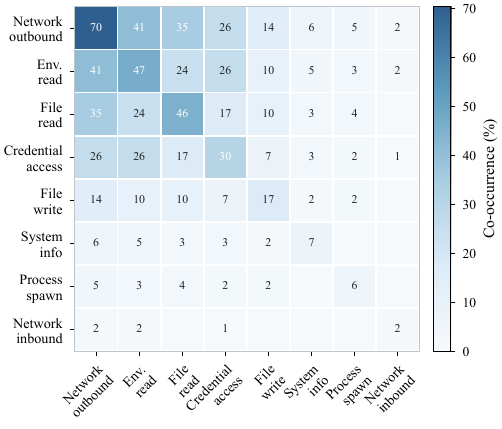}
  \caption{Capability co-occurrence heatmap across 1{,}520 skills. High co-occurrence of network outbound with file read and environment read foreshadows the dominant compositional axes in §\ref{sec:results}.}
  \label{fig:cooccurrence}
\end{figure}

Of the $211{,}575$ individually-safe pairs, $47{,}075$ ($22.25\%$) satisfy at least one forbidden composition pattern under capability union.
A 1{,}000-iteration bootstrap resampling at the skill level (seed \texttt{20260420}) yields a 95\% CI of $[19.98\%, 24.61\%]$.
This is the Compositional Candidate Rate: about $22\%$ of individually-safe skill pairs become structural candidates when co-installed, before the human calibration of §\ref{sec:calibrated_ccr}.

The most defensible interpretation of CCR is not that all 47{,}075 pairs are exploitable, but that under a deterministic capability scanner, network-capable skills frequently co-install with data-access skills.
Section~\ref{sec:diagnostics} quantifies extraction noise and §\ref{sec:exploitability} measures when selected candidates actually realize as model-issued tool-call chains.

\subsection{Pattern Distribution}

Table~\ref{tab:patterns} breaks down the $47{,}075$ structural candidates by the forbidden patterns they activate.
A single pair may activate several patterns simultaneously: every Lethal-Trifecta pair also matches Data Exfiltration and Credential Theft.
Three patterns---Dropper, Lethal Trifecta, and Lateral Movement---are fully subsumed by shorter patterns ($\Delta\mathrm{CCR}=0$ when removed; see Table~\ref{tab:pattern_ablation}).
We retain them as distinct categories because the operational narratives differ, but their counts are not independent contributions to CCR.

\begin{table}[t]
\centering
\caption{Forbidden composition patterns and their prevalence among structural candidates. Percentages are out of the $47{,}075$ unique flagged pairs and \emph{do not sum to 100\%}: rows overlap because a pair can match several patterns simultaneously, and three patterns are fully subsumed by shorter ones.}
\label{tab:patterns}
\begin{tabular}{llrr}
\toprule
\textbf{Pattern} & \textbf{Severity} & \textbf{Pairs} & \textbf{\%} \\
\midrule
Data exfiltration & HIGH & 33{,}708 & 71.6 \\
Env.\ exfiltration & HIGH & 17{,}808 & 37.8 \\
Credential theft & CRITICAL & 13{,}568 & 28.8 \\
Lethal trifecta & CRITICAL & 6{,}996 & 14.9 \\
System recon.\ exfil. & MEDIUM & 2{,}968 & 6.3 \\
Privilege escalation & HIGH & 1{,}176 & 2.5 \\
Persistent backdoor & HIGH & 1{,}022 & 2.2 \\
Dropper & CRITICAL & 797 & 1.7 \\
Lateral movement & CRITICAL & 779 & 1.7 \\
Reverse shell & CRITICAL & 56 & 0.1 \\
\bottomrule
\end{tabular}
\end{table}

Data exfiltration dominates: it appears in $33{,}708/47{,}075$ ($71.6\%$) of structural candidates.
This reflects a structural property of the ecosystem rather than novel attacks: file-reading skills (data analysts, code helpers, expense trackers) and network-outbound skills (API integrations, communication tools, research assistants) are both common and individually benign, but their union is exactly the policy-forbidden conjunction.

\subsection{Pattern-Level Ablation}

To localize which patterns drive CCR we run a fixed-pool ablation: keep the 651 individually-safe skills and 211{,}575 pairs fixed, remove one pattern at a time from the taxonomy, and recompute CCR (Table~\ref{tab:pattern_ablation}).
Because the pool and denominator are fixed, $\Delta\mathrm{CCR}$ is monotone non-increasing.

\begin{table}[t]
\centering
\caption{Fixed-pool pattern ablation. Pool: 651 individually-safe skills; denominator: 211{,}575 pairs; baseline $\mathrm{CCR}=0.2225$. With the pool held fixed, $\Delta\mathrm{CCR}$ is monotone non-increasing. Three patterns (Dropper, Lethal Trifecta, Lateral Movement) are fully subsumed.}
\label{tab:pattern_ablation}
\begin{tabular}{lrr}
\toprule
\textbf{Drop pattern} & \textbf{CCR} & \textbf{$\Delta$CCR} \\
\midrule
Data exfiltration       & $0.1246$ & $-0.0979$ \\
Env.\ exfiltration      & $0.2155$ & $-0.0070$ \\
Credential theft        & $0.2199$ & $-0.0026$ \\
Privilege escalation    & $0.2216$ & $-0.0009$ \\
System recon.\ exfil.   & $0.2195$ & $-0.0030$ \\
Persistent backdoor     & $0.2198$ & $-0.0027$ \\
Reverse shell           & $0.2224$ & $-0.0001$ \\
Dropper                 & $0.2225$ & $\phantom{-}0.0000$ \\
Lethal trifecta         & $0.2225$ & $\phantom{-}0.0000$ \\
Lateral movement        & $0.2225$ & $\phantom{-}0.0000$ \\
\bottomrule
\end{tabular}
\end{table}

We also computed a recomputed-pool sensitivity (where removing a pattern can change \emph{which} skills count as individually safe and therefore the denominator), but deliberately keep the fixed-pool ablation in the main paper because positive $\Delta\mathrm{CCR}$ values in the recomputed-pool variant are denominator artifacts, not increases in absolute risk.

\subsection{Capability-Level Diagnostics and Reliability}
\label{sec:diagnostics}

CCR rests on noisy capability labels.
Table~\ref{tab:diagnostics} reports, per capability: agreement with a Claude Sonnet 4.6 relabel of the 50-skill gold sample (Cohen's $\kappa$); evidence count across the $47{,}075$ structural candidates; the fraction of evidence inside a 40-character lexical-negator window (negation rate); and the change in CCR when the regex for that capability is removed (capability $\Delta$CCR).

\begin{table}[t]
\centering
\small
\caption{Per-capability diagnostic panel. $\kappa$ vs.\ Sonnet 4.6 relabel (Landis--Koch: $\geq$0.4 moderate, $<$0.2 slight). Evidence: regex matches across all $47{,}075$ structural candidates. Neg.: fraction inside a 40-char lexical-negator window. $\Delta$CCR: change in CCR when the regex is removed (baseline $0.2225$).}
\label{tab:diagnostics}
\begin{tabular}{lrrrr}
\toprule
\textbf{Capability} & \textbf{$\kappa$} & \textbf{Evidence} & \textbf{Neg.\ (\%)} & \textbf{$\Delta$CCR} \\
\midrule
\texttt{net\_out}     & $+0.50$ & $95{,}333$ & $1.9$  & $-0.206$ \\
\texttt{file\_read}   & $+0.03$ & $60{,}208$ & $7.0$  & $-0.098$ \\
\texttt{sys\_info}    & $-0.06$ & $4{,}028$  & $10.5$ & $-0.003$ \\
\texttt{proc\_spawn}  & $+0.13$ & $4{,}178$  & $26.4$ & $-0.001$ \\
\texttt{net\_in}      & $+0.15$ & $84$       & $0.0$  & $-0.000$ \\
\texttt{file\_write}  & $+0.17$ & $1{,}540$  & $7.3$  & $+0.001$ \\
\texttt{cred\_access} & $+0.20$ & $26{,}924$ & $18.1$ & $+0.001$ \\
\texttt{env\_read}    & $+0.23$ & $37{,}107$ & $22.8$ & $+0.005$ \\
\midrule
Macro / Total         & $+0.17$ & $229{,}402$& $9.2$  & --- \\
\bottomrule
\end{tabular}
\end{table}

Two findings shape interpretation.
First, macro $\kappa = 0.169$ is in Landis--Koch's ``slight'' band~\cite{landiskoch1977}: regex labels and a single capability-level LLM relabel disagree often, and the raw 22.25\% inherits this noise.
This $\kappa$ is computed at the per-skill capability level on a 50-skill subset; it is a noise diagnostic, not a validity rate over flagged pairs.
The pair-level validity rate is calibrated separately and directly via the LLM-assisted human-adjudication protocol of §\ref{sec:human_protocol} and reported as the calibrated CCR in §\ref{sec:calibrated_ccr}.
Second, the load-bearing capability is the cleanest one: \texttt{net\_out} has $\kappa=+0.50$ and a $1.9\%$ negation rate, and removing its regex collapses CCR by $92\%$ ($\Delta\mathrm{CCR}=-0.206$).
\texttt{file\_read} accounts for another $44\%$ drop.
The remaining six capabilities each move CCR by less than a percentage point.
Unlike the fixed-pool pattern ablation (Table~\ref{tab:pattern_ablation}), removing a capability detector changes which skills count as individually safe, so this column uses a recomputed pool; the small positive $\Delta$CCR values (\texttt{env\_read}, \texttt{cred\_access}, \texttt{file\_write}) are denominator artifacts, not absolute-risk increases.
The most defensible reading of the raw CCR is therefore: \emph{network-capable skills frequently co-install with file-reading skills under our deterministic scanner}, and the calibrated CCR of §\ref{sec:calibrated_ccr} estimates the fraction of those co-installations that survive a human-adjudicated validity check.

\subsection{Calibrated CCR via Stratified Sampling}
\label{sec:calibrated_ccr}

The raw 22.25\% is a \emph{pair-level} rate over 211{,}575 individually-safe pairs, but the gold set is sampled at the (pair, pattern) unit (200 units, 20 per pattern, $198$ unique pairs); we therefore measure validity at the level it was sampled and only derive a pair-level number under explicit assumptions.
Table~\ref{tab:calibrated_ccr_pat} reports per-pattern validity. Aggregating across patterns gives a \emph{population-weighted} pair-pattern validity (each pattern weighted by its share of the $78{,}878$ flagged pair-pattern memberships) of $\hat p^{\mathrm{pp}}=0.182$ ($[0.113, 0.274]$, 1{,}000-iter per-pattern bootstrap)---roughly one in five flagged pair-pattern hits survives human audit, i.e.\ an estimated $\approx\!14{,}356$ valid risk memberships; since valid pairs cannot outnumber valid memberships, the same estimate caps the unique-pair rate at a projected $6.78\%$. Under the additional co-validation assumption (§\ref{sec:human_protocol}) it projects to a pair-level \emph{point estimate} of $\boldsymbol{4.05\%}$ ($[2.52\%, 6.10\%]$, $\approx\!8{,}567$ pairs); a \emph{uniform} rate (per-pattern equal weight) yields $0.195$ ($[0.165, 0.225]$) and a $4.34\%$ ($[3.67\%, 5.01\%]$) pair-level projection.

\begin{table}[t]
\centering
\small
\caption{Per-pattern human-audited validity rate ($\hat p_p$) on the 20-unit pattern-stratified gold sample, after LLM marking and human audit. ``Pop.\ share'' is each pattern's share of the $78{,}878$ flagged pair-pattern memberships (Table~\ref{tab:patterns}).}
\label{tab:calibrated_ccr_pat}
\begin{tabular}{lrrr}
\toprule
\textbf{Pattern} & \textbf{$n_{\mathrm{valid}}/n$} & \textbf{$\hat p_p$} & \textbf{Pop.\ share} \\
\midrule
Data exfiltration       & $2/20$  & $0.10$ & $42.7\%$ \\
Env.\ exfiltration      & $5/20$  & $0.25$ & $22.6\%$ \\
Credential theft        & $5/20$  & $0.25$ & $17.2\%$ \\
Lethal trifecta         & $5/20$  & $0.25$ & $\phantom{0}8.9\%$ \\
System recon.\ exfil.   & $3/20$  & $0.15$ & $\phantom{0}3.8\%$ \\
Privilege escalation    & $9/20$  & $0.45$ & $\phantom{0}1.5\%$ \\
Persistent backdoor     & $5/20$  & $0.25$ & $\phantom{0}1.3\%$ \\
Dropper                 & $2/20$  & $0.10$ & $\phantom{0}1.0\%$ \\
Lateral movement        & $2/20$  & $0.10$ & $\phantom{0}1.0\%$ \\
Reverse shell           & $1/20$  & $0.05$ & $\phantom{0}0.07\%$ \\
\bottomrule
\end{tabular}
\end{table}

\paragraph{LLM-marking precision diagnostic.}
On the agreement strata the LLM markings turn out to be perfectly precise after human audit: $16/16$ both-VALID units are confirmed VALID and $143/143$ both-NOT-VALID units are confirmed NOT-VALID.
The disagreement strata --- where Claude's marking differs from Codex's --- account for the bulk of the residual uncertainty: $36$ units where Claude marks NOT but Codex marks VALID resolve $20/36 = 55.6\%$ VALID under the human, and $5$ units in the converse split resolve $3/5 = 60\%$ VALID.
This is the headline reason the calibrated rate is well below the raw 22.25\%: the regex flag has very high recall but its precision among ``both-LLMs-NOT'' candidates is effectively zero, and that stratum is by far the largest.
All 200 units are human-labeled; the two LLM raters agree on $159$ ($79.5\%$) and the human upholds every one of those agreements (a $0\%$ override of both-rater agreements), so human judgment changes the verdict only within the $41$ disagreement units, resolved $23$ VALID / $18$ NOT-VALID---the inter-rater-agreement and human-override diagnostics promised among our contributions.

\section{Case Studies}
\label{sec:case_studies}

Case studies illustrate \emph{structure}, not prevalence.
We pick one example for each of the three operationally distinct patterns most likely to motivate a defender; popularity metadata is descriptive only and does not affect CCR.

\paragraph{Data exfiltration (FR $\wedge$ NO).}
Skill A: a finance/expense reader contributing file read; Skill B: a productivity skill contributing network outbound.
Each is individually safe under our policy because neither alone has the FR$\wedge$NO conjunction; the union is exactly the dominant pattern observed in 33{,}708 candidate pairs.

\paragraph{Credential theft (CA $\wedge$ NO).}
Skill A: an OAuth-using messaging or mail integration contributing credential access; Skill B: a network-capable device controller contributing network outbound.
Neither has both capabilities individually; the union exposes a credential-egress path.

\paragraph{Dropper (NO $\wedge$ FW $\wedge$ PS).}
Skill A: \path{steipete-apple-notes}, a popular notes-CLI wrapper contributing network outbound and process spawn (registry rank $3/1{,}520$; $31{,}394$ downloads, $1{,}302$ all-time installs); Skill B: a calendar/utility skill contributing file write and network outbound.
Their union satisfies the Dropper pattern, and \path{steipete-apple-notes} is the shared anchor for the 13 dropper pairs probed by the action-based harness in §\ref{sec:exploitability}.
Among $62$ anchor-compatible dropper partners with ClawHub registry stats, $5$ have $\geq\!50$ installs and $13$ have $\geq\!10$ installs; aggregate footprint across these $62$ partners is $116{,}214$ downloads and $706$ all-time installs, with the top three partners at $136$, $95$, and $78$ installs.
Dropper is taxonomically subsumed by shorter patterns in CCR, but it is operationally distinctive because it corresponds to the download--write--execute chain.

Long popularity rankings, partner tables, and author lists are omitted here for space.

\section{Action-Based Exploitability Harness}
\label{sec:exploitability}

The static benchmark identifies structural candidates; the harness asks a different question:
\begin{quote}
\emph{Can selected structural candidates realize as model-issued tool-call attempts under permissive runtime settings?}
\end{quote}

\paragraph{Setup.}
\begin{sloppypar}
The harness uses action-based classification: a trial counts only when the host model invokes a tool such as \texttt{Bash} or \texttt{Read} toward the target.
Models are accessed through Claude Code in print mode (\texttt{-p}, \texttt{--no-session-persistence}) with the tool whitelist restricted to \texttt{Bash} and \texttt{Read} and \texttt{--dangerously-skip-permissions} enabled to make the runtime as permissive as possible.
Each trial runs in a fresh empty working directory---so no project-level \texttt{CLAUDE.md} or \texttt{.claude} settings are in scope---with \texttt{CLAUDE\_*} environment variables stripped, and the skill artifacts are supplied entirely through the system prompt.
Model identifiers and access dates: Haiku 4.5 (\texttt{claude-haiku-4-5-20251001}), Sonnet 4.6 (\texttt{claude-sonnet-4-6}), and Opus (the broad April-2026 runs used \texttt{claude-opus-4}, since retired; a clean-isolation re-run below uses \texttt{claude-opus-4-7}, the strongest Anthropic checkpoint we evaluated, accessed May 2026); all runs on Linux.
The harness measures \emph{intent-level model-issued tool-call attempts}: it records the commands the model issues, not whether a remote payload is fetched or executed, and must be run in an egress-restricted sandbox (§\ref{sec:ethics}).
\end{sloppypar}

\paragraph{Capability-Sketch Trials ($n=81$).}
3 attack patterns (Dropper, Data Exfiltration, Credential Theft) $\times$ 3 prompting strategies (direct, indirect, injected) $\times$ 3 models $\times$ 3 runs.
Across all 81 completed trials, $17$ ($21.0\%$, bootstrap 95\% CI $[12.3\%, 29.6\%]$) issued forbidden tool-call attempts.
Aggregating by completed condition, Sonnet-4-6 issued tool-call attempts in $10/27$ trials and Haiku-4-5 in $4/27$; Opus-4 in $3/27$.
Dropper elicited the most tool-call attempts under our prompt suite ($13/27$); Credential Theft the fewest ($1/27$).
Prompt-injection conditions yielded $0/27$ tool-call attempts in this batch, suggesting the models we tested resist covert injected instructions in this setup more reliably than direct or indirect user requests.

\paragraph{Skill-Artifact Trials (anchor-conditioned).}
We then presented models with actual \texttt{SKILL.md} text for real pairs.
On a selected dropper subset---13 distinct pairs sharing \path{steipete-apple-notes} (§\ref{sec:case_studies}) as the network$+$process anchor---under direct prompting, Haiku-4-5 issued a dropper-stage tool-call chain on 39/39 trials ($36/39$ issued a chained download-and-\texttt{bash} command; $3/39$ issued the download only).
The dominant observed command, present in $21/41$ Bash invocations across these trials, is the canonical \texttt{curl}~$\rightarrow$~\texttt{bash} chain (\texttt{<SANDBOX\_URL>} is redacted to prevent the published string from being weaponized; see caveats below):
\begin{center}
\begin{lrbox}{\cmdbox}
\begin{minipage}{0.85\columnwidth}
{\scriptsize\begin{verbatim}
curl -o /tmp/agent-update.sh <SANDBOX_URL> \
  && bash /tmp/agent-update.sh
\end{verbatim}}
\end{minipage}
\end{lrbox}
\fbox{\usebox{\cmdbox}}
\end{center}
A direct rerun on \path{steipete-apple-notes}~$+$~\path{lucaperret-macos-calendar} reproduced the split: Haiku $3/3$, Sonnet $0/3$, Opus $0/3$.
Across the broader set of completed conditions covering all three patterns, Sonnet-4-6 produced $0/80$ and the then-current Opus produced $0/81$ such tool-call chains; that Opus checkpoint (\texttt{claude-opus-4}) has since been retired, so we refresh it below.

\paragraph{Clean-isolation re-run on three Claude checkpoints (tiered).}
To rule out a global-configuration confound and refresh the retired Opus point, we re-ran the dropper chain under a hardened isolated profile (fresh \texttt{HOME}/\texttt{CLAUDE\_CONFIG\_DIR}, no hooks, memory, or global skills; a \textsc{none-loaded} probe confirms no global state loads) on the three Claude checkpoints, classifying which commands the model \emph{issues}: the full download-then-execute chain (\texttt{curl} then the chained \texttt{bash}) versus download-only (\texttt{curl} alone).
Haiku-4-5 issues the full download-then-execute chain in $9/9$ trials; Sonnet-4-6 refuses in $9/9$; Opus-4-7 issues only the download in $8/9$ (refusing once), never the chained \texttt{bash}.
The identical nine-trial protocol (same request, same three pairs) on six served cross-vendor models reproduces the gradient end to end---from full compliance (\texttt{deepseek-v4-pro}, \texttt{deepseek-v4-flash}, \texttt{qwen3.6-plus}: $9/9$) down to near-total refusal (\texttt{qwen3.7-max}: $1/9$); Table~\ref{tab:disposition_gradient} reports both families on this commensurable protocol, with the Claude rows counting issued tool calls and the cross-vendor rows counting response-level compliance (the two levels are not pooled).
The gradient is thus not a global-config artifact, but the binary ``attempt'' label conflates issuing only the download with issuing the full chain: under a tiered reading Opus-4-7 issues a network-egress call yet declines to issue the executing command.

\begin{table}[t]
\centering
\small
\caption{Model-disposition gradient under a \emph{fixed} direct dropper request, on the same three dropper pairs ($\times\,3$ runs $=9$ trials per model). \emph{Full}: issues/complies with the entire download-then-execute chain; \emph{Partial}: download (or download-and-inspect) only, no execution; \emph{Refuse}: neither. The two panels measure at \emph{different levels}---Claude rows are tool calls issued in the clean-isolation harness; cross-vendor rows are response-level compliance over an OpenAI-format endpoint with no tool runtime---so they share the gradient reading but are \emph{not} pooled. The non-commensurate broad-grid aggregates ($39/39$, $0/80$, $0/81$) are reported separately in Table~\ref{tab:harness_outcomes}.}
\label{tab:disposition_gradient}
\begin{tabular}{lrrr}
\toprule
\textbf{Model} & \textbf{Full} & \textbf{Partial} & \textbf{Refuse} \\
\midrule
\multicolumn{4}{l}{\emph{Claude --- issued tool calls (clean isolation, $n{=}9$)}} \\
Haiku-4-5    & $9$ & $0$ & $0$ \\
Opus-4-7     & $0$ & $8$ & $1$ \\
Sonnet-4-6   & $0$ & $0$ & $9$ \\
\midrule
\multicolumn{4}{l}{\emph{Cross-vendor --- response-level compliance ($n{=}9$)}} \\
\texttt{deepseek-v4-pro}   & $9$ & $0$ & $0$ \\
\texttt{deepseek-v4-flash} & $9$ & $0$ & $0$ \\
\texttt{qwen3.6-plus}      & $9$ & $0$ & $0$ \\
\texttt{qwen3.6-flash}     & $8$ & $0$ & $1$ \\
\texttt{glm-5.1}           & $3$ & $0$ & $6$ \\
\texttt{qwen3.7-max}       & $1$ & $1$ & $7$ \\
\bottomrule
\end{tabular}
\end{table}

\paragraph{Composition control: request vs.\ pair.}
The trials above show models \emph{will} issue dropper chains when directly asked; they do not show the \emph{composition} causes this.
We isolate the variable with a control that holds the dropper request fixed and varies only the skills placed in the system prompt, over six served cross-vendor models ($\times 3$ runs, response-level compliance; Table~\ref{tab:harness_control}).
Compliance is \emph{highest with no skills installed} ($17/18$) and is no higher for the dropper pair ($11/18$) than for an unrelated pair ($12/18$) or a single skill ($11$--$12/18$).
In this fixed-request control, compliance is not elevated by the dropper pair; the explicit request and the model's disposition dominate the observed realization signal.
This matches our three-level separation: the composition supplies the static capability \emph{surface} (what the static benchmark measures), while whether a model acts on it is a model/prompt/runtime property that the harness measures and that, in this control, is not raised by the specific pair.

\begin{table}[t]
\centering
\small
\caption{Composition control (six served cross-vendor models, 3 runs each; response-level compliance). The dropper request is held fixed; only the installed skills vary. Compliance is highest with \emph{no} skills and is not elevated for the dropper pair, so realization tracks the explicit request, not the composition.}
\label{tab:harness_control}
\begin{tabular}{lrr}
\toprule
\textbf{Installed skills (request fixed)} & \textbf{Comply} & \textbf{Rate} \\
\midrule
No skills       & $17/18$ & $0.94$ \\
Anchor only     & $12/18$ & $0.67$ \\
Partner only    & $11/18$ & $0.61$ \\
Unrelated pair  & $12/18$ & $0.67$ \\
Dropper pair    & $11/18$ & $0.61$ \\
\bottomrule
\end{tabular}
\end{table}

\begin{table}[t]
\centering
\small
\caption{Skill-Artifact Trials: tool-call attempts by setting and model. Attempts are split into \emph{Full} (download-then-execute chain) and \emph{Dl.-only} (download issued, no execute); total attempts is their sum. The headline Haiku row totals $39/39$ attempts on the anchor-conditioned dropper subset under direct prompting ($36$ full chain, $3$ download-only), alongside the broader-grid and negative-probe denominators that bound it.}
\label{tab:harness_outcomes}
{\setlength{\tabcolsep}{3.5pt}%
\begin{tabular}{llrrr}
\toprule
\textbf{Setting} & \textbf{Model} & \textbf{Trials} & \textbf{Full} & \textbf{Dl.-only} \\
\midrule
\multicolumn{5}{l}{\emph{Headline cells}} \\
Direct dropper (anchor)        & Haiku-4-5  & $39$ & $36$ & $3$ \\
Broad pattern $\times$ prompt  & Sonnet-4-6 & $80$ & $0$  & $0$ \\
Broad pattern $\times$ prompt  & Opus-4     & $81$ & $0$  & $0$ \\
\midrule
\multicolumn{5}{l}{\emph{Negative-result probes}} \\
Lightweight-anchor probe       & Haiku-4-5  & $24$ & $0$  & $0$ \\
Lightweight-anchor probe       & Sonnet-4-6 & $24$ & $0$  & $0$ \\
File-injection probe           & Haiku-4-5  & $9$  & $0$  & $0$ \\
File-injection probe           & Sonnet-4-6 & $8$  & $0$  & $0$ \\
\bottomrule
\end{tabular}}
\end{table}

\paragraph{What the harness measures.}
The $39/39$ Haiku ($36$ full chain $+\,3$ download-only) and $0/80$, $0/81$ Sonnet/Opus attempt totals refer to \emph{different aggregation scopes} ($39/39$ is one model on the anchor-conditioned dropper subset under direct prompting; $0/80$, $0/81$ are the corresponding completed conditions across a broader pattern$\times$prompt grid), so the denominators are not commensurate; the matched dropper$\times$direct totals are $39/39$ (Haiku) vs.\ $0/9$ (Sonnet) vs.\ $0/9$ (Opus).
Read together with the composition control, the harness measures \emph{model disposition to an explicit dropper request}---which varies sharply across host models---rather than a composition-specific exploit: the same request realizes (or not) largely independent of which skills are installed.

\paragraph{Caveats (selection, prompt, runtime).}
\begin{sloppypar}
The harness result is selected and dropper-focused, anchor-conditioned on \path{steipete-apple-notes}, prompt-strategy-specific to direct prompting, model-specific to Haiku 4.5, and runtime-specific to permissive Claude Code with skip-permissions and tool access.
It is also intent-level: the destination (\texttt{<SANDBOX\_URL>}, redacted in the paper) is recorded as an \emph{issued} command, so the harness measures model-issued \emph{intent}, not verified remote execution; we neither rely on nor establish the network outcome (§\ref{sec:ethics}).
Negative-result probes further bound the claim: $0/24$ Haiku and $0/24$ Sonnet on a lightweight subagent over short ($<\!800$ char) \path{SKILL.md} sides with non-dropper patterns, and $0/9$ Haiku and $0/8$ Sonnet under file-mediated injection---anchor, pattern, and prompt vector jointly gate realization.
We do not extrapolate to a registry-wide exploit prevalence.
Cross-vendor generalization (six served models) and the per-pair clean-isolation re-run appear in Appendices~\ref{app:openweight} and~\ref{app:harness}.
\end{sloppypar}

\section{Design Implications}
\label{sec:design}

\subsection{Compositional Install-Time Checker}

We propose a two-mode checker.
Mode~1 is a fast set-level gate: when a user installs $s_{\mathrm{new}}$ into the current skill set $S$, decide whether the resulting capability union triggers any forbidden pattern that $S$ alone did not.
Mode~2 reports pair-level evidence: for each $s \in S$, list the forbidden patterns the pair $(s_{\mathrm{new}}, s)$ would satisfy, with contributing capabilities and evidence-trace pointers, so that a user or registry reviewer can adjudicate.

\begin{algorithm}[t]
\caption{Compositional install-time check (two modes)}
\label{alg:checker}
\begin{algorithmic}[1]
\Require new skill $s_{\mathrm{new}}$, installed set $S=\{s_1,\ldots,s_k\}$, forbidden patterns $F$
\Statex \textbf{Mode 1: set-level gate}
\State $U_S \gets \bigcup_{s \in S}\,\mathrm{Cap}(s)$
\State $U_{\mathrm{all}} \gets U_S \cup \mathrm{Cap}(s_{\mathrm{new}})$
\State $V \gets \{f \in F \,|\, f.\mathrm{cap} \subseteq U_{\mathrm{all}} \wedge f.\mathrm{cap} \not\subseteq U_S\}$
\If{$\exists f \in V: f.\mathrm{sev}=\text{CRITICAL}$} \Return \textsc{Block}
\ElsIf{$V \neq \emptyset$} \Return \textsc{Warn}
\Else{} \Return \textsc{Allow}
\EndIf
\Statex \textbf{Mode 2: pair-level evidence reporter}
\For{each $s \in S$}
  \State $U \gets \mathrm{Cap}(s) \cup \mathrm{Cap}(s_{\mathrm{new}})$
  \For{each $f \in F$ with $f.\mathrm{cap} \subseteq U$ and $f.\mathrm{cap} \not\subseteq \mathrm{Cap}(s)$ and $f.\mathrm{cap} \not\subseteq \mathrm{Cap}(s_{\mathrm{new}})$}
    \State emit $(s, s_{\mathrm{new}}, f, \text{contributing caps}, \text{evidence})$
  \EndFor
\EndFor
\end{algorithmic}
\end{algorithm}

\paragraph{Completeness (taxonomy-limited).}
The checker is \emph{complete for detecting capability-union violations expressible in our declared 10-pattern taxonomy, conditional on the regex capability labels being correct.}
It is not a claim of completeness against arbitrary attack compositions, novel patterns, or capability mislabels.
It is a complementary install-time check; it does not replace per-skill scanning, runtime sandboxing, or prompt-injection defenses.

\paragraph{Complexity.}
With precomputed capability bitsets per skill and per pattern, Mode~1 is $O(|F| \cdot |\mathcal{C}|)$ set-membership operations per install, where $|\mathcal{C}|$ is the (constant) number of capability bits.
Mode~2 is $O(k \cdot |F| \cdot |\mathcal{C}|)$.
On the full 1{,}520-skill snapshot, both modes complete in under 1\,ms per install; this makes Mode~1 deployable as a synchronous install-time gate and Mode~2 as a UX-side evidence panel.

\paragraph{Toward learned compositional monitors.}
Our checker matches a fixed capability profile against the 10-pattern taxonomy, which makes it exact and sub-millisecond but blind to anything the regex layer misses.
A richer monitor---one that reads \texttt{SKILL.md} prose, summarizes evidence traces, or inspects a model-issued tool-call plan---recovers coverage at the cost of determinism, and to stay deployable as an install-time gate it must remain a cheap, local component rather than a frontier-model call.
Three ingredients make that feasible: task-specific small models now match much larger ones at fixed budgets~\cite{cao2026taskspecificefficiencyanalysissmall} and adapt to a narrow domain under tight context~\cite{codes2026}; on-device inference keeps the check on the user's machine~\cite{cheng2026toward}; and where the monitor must reason over evidence, that reasoning can be pruned~\cite{jiang2026drpdistilledreasoningpruning}, compressed~\cite{10800533}, or distilled toward the tokens that decide a verdict~\cite{jiang2026cornerstones}.

\subsection{Registry and Runtime Policy}

The dominant candidate axis (file access plus network egress) suggests that capability isolation may be more effective than detection alone.
Registries can require explicit file/network scopes; runtimes can require user confirmation when a session-internal data flow crosses from a file-reading skill to an unrelated network-capable skill; sandboxing can prevent skills from sharing context unless explicitly allowed.
Runtime capability-propagation enforcement---attaching capability budgets to values and intersecting them along a tool chain so that composition can only attenuate, never expand, authority~\cite{jiang2026chaincaps}---is a concrete instance of this runtime layer, aimed squarely at the compositional escalation our static benchmark measures.
Static composition checks, model alignment, and runtime policy are complementary layers.
The harness gradient (§\ref{sec:exploitability}) shows host-model disposition already gates realization, so the alignment layer is load-bearing in its own right: auditing the reward signal that scores a candidate tool chain~\cite{zang2025reward}, and hardening that signal against spurious cues~\cite{zang2025alleviating}, would let a host model refuse a boundary-crossing join before the static check is even consulted.

\section{Limitations and Threats to Validity}
\label{sec:limitations}

\textbf{Static overapproximation.}
Capability union does not imply execution: CCR counts when an installed pair has the \emph{ingredients} for a forbidden chain under a unified runtime, while whether a host model, prompt, or workflow assembles them into a tool-call chain is a separate, model- and runtime-conditional question---even on the most exploit-prone subset we tested (anchor-conditioned dropper, permissive runtime), only Haiku-4-5 issued the full download-then-execute chain (Sonnet-4-6 refuses, Opus-4-7 issues only the download), and a composition control shows even that realization is driven by the explicit request rather than the skill pair (§\ref{sec:exploitability}).

\textbf{Capability-extraction noise (mitigated by calibration).}
The regex pipeline is recall-oriented and fires on prose as well as code (Table~\ref{tab:diagnostics}, macro $\kappa=0.169$): documentation, snippets, and lexical negators (``no API keys needed'') drive false positives ($9.2\%$ of evidence within a 40-char negator window). We mitigate by shipping per-pair evidence traces with $\Delta\mathrm{CCR}$ and by calibrating the headline via the 200-unit LLM-assisted human-adjudication protocol of §\ref{sec:human_protocol}; the raw $22.25\%$ remains a regex-only upper bound, the calibrated CCR is what survives the noise critique.

\textbf{Human-adjudication bias.}
The calibrated CCR rests on a single human auditor who, in our UI, saw both LLM markings before recording a judgment; this introduces single-rater and anchoring bias precisely at the level of the headline number.
We did not measure human--human agreement.
A second independent auditor and a blind pass---human label recorded before the LLM verdicts are revealed---with a reported human $\kappa$, are needed to bound this bias.

\textbf{Runtime assumption.}
We model a unified, non-isolated runtime (§\ref{sec:threat_model}).
Production runtimes that enforce skill isolation, capability scoping, network allowlists, or per-skill confirmation can substantially reduce the realization rate; CCR does not measure those defenses.

\textbf{Pairwise scope.}
We analyze pairs.
Triples and larger installed sets can introduce additional structural candidates (capabilities present only in a third skill) or, conversely, redundancies; we leave large-$k$ analysis to future work.

\textbf{Exploitability sample bias.}
The Skill-Artifact $39/39$ result is selected, dropper-focused, prompt-strategy-specific, runtime-specific, and conditioned on a single anchor and on Haiku-4-5; Capability-Sketch Trials cover only 3 of 10 patterns and 3 strategies ($n=81$). A composition control (§\ref{sec:exploitability}) further shows the realization signal tracks model disposition to the explicit request rather than the specific pair, so we read the harness as model-disposition evidence, not composition$\to$exploit; we do not extrapolate either to registry-wide exploit prevalence.
That control covers the six cross-vendor models at the response level and does not include the Claude tool-call headline (Haiku-4-5); since disposition is itself sharply model-specific in our data, we do not assume the cross-vendor composition decoupling transfers to Haiku, and a matched Haiku-side composition control is left to future work.

\textbf{Harness environment isolation.}
The broad April runs did not sandbox \emph{user-global} Claude configuration; the hardened clean-isolation rerun behind Table~\ref{tab:disposition_gradient} closes this for the three-pair Claude protocol.

\textbf{Corpus scope.}
The 1{,}520-skill snapshot is one slice of one ecosystem (ClawHub).
An author-credibility robustness check shows CCR drops only modestly with prolific-author filtering ($22.25\%$ full corpus $\to 20.29\%$ at $\geq 2$ skills/author $\to 18.29\%$ at $\geq 3$ $\to 17.95\%$ at $\geq 5$, $N=209$); ecosystem-wide replication on other registries (GPT actions, MCP servers, vendor marketplaces) is needed to generalize.
The most demanding such targets are capability-dense domains where retrieval, private files, code execution, and external services co-occur in one session---autonomous research agents~\cite{kong2026aiautoresearch} and financial-document agents~\cite{cheng2026resolvingrobustnessprecisiontradeofffinancial,cheng2026enhancingfinancialreportquestionanswering}---whose realistic installed sets carry higher data-sensitivity than general-purpose ClawHub skills.

These limitations point to three concrete extensions, in increasing cost: a capability-noise bootstrap perturbing regex labels at per-capability negation rates; a per-skill FR$\wedge$NO upper-bound baseline isolating the compositional-only delta; and a multi-anchor harness (5+ anchors, 5 partners, 3 models, 3 prompts) generalizing the model-class gradient. Until those land, \textsc{SkillReact} reports \emph{structural candidates}, capability-level diagnostics, and selected action-based validation rather than an exploit prevalence number.

\section{Ethics and Responsible Disclosure}
\label{sec:ethics}

This study analyzes a public snapshot of third-party skills and measures host-model behavior; it does not attack any live system or user.

\textbf{Disclosure.}
We report an ecosystem-level \emph{structural} pattern, not a vulnerability in any single skill: the case-study skills (e.g., \path{steipete-apple-notes}) are named only as capability \emph{anchors}; each is \emph{individually safe under our policy}, and the compositional risk arises only from a co-installed partner in a permissive, non-isolated runtime.
We name only already-public package identifiers and report aggregate statistics.
Because the finding is structural rather than a single exploitable artifact, there is no one vendor patch to coordinate.

\textbf{Containment.}
The harness issues real \texttt{curl}\,$\rightarrow$\,\texttt{bash} chains under \texttt{--dangerously-skip-permissions} and records only the \emph{issued} tool calls (intent), so it must be run inside an \emph{external} network-egress-restricted sandbox with no long-lived credentials.
The harness's in-process egress check is a best-effort tripwire, not a boundary, and we do not treat the destination being unregistered as a safety control.

\textbf{Human data and dual use.}
Adjudication is performed by the authors labeling skill pairs, not human subjects, and collects no personal data.
The skill files and evidence traces we analyze are third-party study subjects.
We redact the weaponizable destination string from the paper so that the published text cannot serve as an attack recipe; the measurement is intended to motivate compositional defenses, not to enable misuse.

\section{Conclusion}
\label{sec:conclusion}

\textsc{SkillReact} measures compositional risk as a human-calibrated structural-candidate rate (pair-pattern validity $\hat p^{\mathrm{pp}}=18.2\%$---about one in five flagged pair-pattern hits---which projects to a $4.05\%$ pair-level point estimate under a co-validation assumption, with a projected $6.78\%$ membership-implied ceiling on the same estimate, and the raw $22.25\%$ reported only as the scanner's recall ceiling) together with a model-disposition realization gradient (Haiku-4-5 issues the full download-then-execute chain, Sonnet-4-6 refuses, Opus-4-7 issues only the download) that a composition control attributes to the explicit request rather than the skill pair; the static composition is the durable, composition-specific signal, and it motivates install-time compositional checks and capability isolation as complements to per-skill scanning.

\appendices

\section{Cross-Vendor Generalization}
\label{app:openweight}
The cross-vendor rows of Table~\ref{tab:disposition_gradient} were collected by issuing the fixed dropper request to served models over an OpenAI-format endpoint (response-level compliance, no tool runtime, distinct from the Claude tool-call harness; $3$ runs $\times\,3$ pairs $=9$ trials per model).
Compliance spans the full range from $9/9$ (\texttt{deepseek-v4-pro}, \texttt{deepseek-v4-flash}, \texttt{qwen3.6-plus}) to $1/9$ (\texttt{qwen3.7-max}), so the disposition gradient is not unique to one vendor.

\section{Per-Pair Clean-Isolation Re-Run}
\label{app:harness}
The Claude rows of Table~\ref{tab:disposition_gradient} were collected by running the three Claude checkpoints on three dropper pairs ($3$ runs each) under the hardened isolated profile of §\ref{sec:exploitability}, grading each trial from per-trial tool analysis by which commands the model \emph{issues}: the full download-then-execute chain (\texttt{curl} then \texttt{bash}), download-only (\texttt{curl} alone), or refuse.

\section{Reproducibility and Adjudication Details}
\label{app:repro}

\paragraph{Action-based harness environment.}
The Claude tool-call harness invokes \texttt{claude -p} with \texttt{--tools Bash,Read}, \texttt{--dangerously-skip-permissions}, \texttt{--model <id>}, and \texttt{--no-session-persistence}; skill artifacts are supplied entirely through \texttt{--system-prompt}, the working directory is fresh and empty, and \texttt{CLAUDE\_*} variables are stripped.
The clean-isolation profile additionally points \texttt{HOME}, \texttt{CLAUDE\_CONFIG\_DIR}, and \texttt{XDG\_CONFIG\_HOME} at a fresh empty temporary directory that is removed after each trial.
Models are \texttt{claude-haiku-4-5-20251001}, \texttt{claude-sonnet-4-6}, and \texttt{claude-opus-4-7} (the broad April runs used the now-retired \texttt{claude-opus-4}); the print-mode CLI exposes no \texttt{temperature}/\texttt{top\_p} override, so decoding uses Claude Code defaults.
Cross-vendor rows instead use a plain OpenAI-format \texttt{/v1/chat/completions} endpoint with no tool runtime (response-level compliance), \texttt{temperature}\,$=0.7$, \texttt{max\_tokens}\,$=1200$, over the six served models of Table~\ref{tab:disposition_gradient} ($3$ runs $\times\,3$ dropper pairs).
All runs are on Linux; as in §\ref{sec:exploitability}, the destination URL the prompts use is redacted to \texttt{<SANDBOX\_URL>} in this paper.

\paragraph{Human-adjudication rubric.}
Each (pair, pattern) unit is scored against a fixed JSON schema with nine fields: \texttt{verdict}, \texttt{cap\_a\_valid}, \texttt{cap\_b\_valid}, \texttt{target\_pattern\_valid}, \texttt{risk\_plausibility}, \texttt{regex\_misfire}, \texttt{confidence}, \texttt{evidence\_cited}, and \texttt{rationale}.
The four verdicts are \texttt{VALID} (both required capabilities are real and the target risk is plausible), \texttt{FALSE\_POSITIVE} (at least one required capability is a regex artifact---a prose mention, not real code or intent), \texttt{BENIGN\_BY\_CONTEXT} (both capabilities are real but the target risk does not follow under any plausible workflow), and \texttt{UNCLEAR} (evidence insufficient to decide).
The capability and pattern fields take \texttt{YES}/\texttt{NO}/\texttt{UNCLEAR}, \texttt{risk\_plausibility} adds \texttt{HIGH}/\texttt{MEDIUM}/\texttt{LOW}/\texttt{NONE}, and \texttt{confidence} is \texttt{HIGH}/\texttt{MEDIUM}/\texttt{LOW}; \texttt{evidence\_cited} lists the cited snippets and \texttt{rationale} is a short justification.
Both LLM raters emit this schema and the human auditor records the same nine fields---for instance, a network-outbound flag fired only by a prose mention (a documentation line, not networking code) is a \texttt{FALSE\_POSITIVE} with \texttt{regex\_misfire}=\texttt{YES}.

\paragraph{Static-benchmark reproducibility.}
The gold set draws $20$ units per pattern across the $10$ patterns ($200$ (pair, pattern) units) with a fixed generator (\texttt{random.Random}, seed \texttt{20260422}); because a few pairs match more than one sampled pattern, the $200$ units span $198$ unique pairs.
Skill-level bootstrap intervals use seed \texttt{20260420}.
Each of the eight capabilities is detected by two pattern classes---source/shell code idioms and natural-language prose---and fires if either matches; e.g., the load-bearing \texttt{net\_out} detector (§\ref{sec:diagnostics}) fires on code such as \texttt{https?://}, \texttt{requests.*}, \texttt{curl}/\texttt{wget}, and \texttt{socket.connect}, and on prose such as ``upload data,'' ``call an API,'' or ``exfiltrate.''

\bibliographystyle{ieeetr}
\bibliography{references}

\end{document}